\documentstyle[12pt]{article}
\begin{document}

\begin{titlepage}
\vspace{2.5cm}
\begin{centering}
{\Large{\bf The partition function\\}}
{\Large{\bf of the unit interval}}\\
\bigskip\bigskip
Ram\'on Mendoza$^{\ddagger}$ and Fernando Moraes$^{\star}$\\
{\em Departamento de Matem\'atica$^{\ddagger}$ and Departamento de F\'{\i}sica$^{\star}$}\\
{\em Universidade Federal de Pernambuco}\\
{\em 50670-901 Recife, PE, Brazil}\\

\end{centering}
\vspace{1.5cm}
\begin{abstract}
We outline a method of calculation of partition functions of orientable
manifolds with fluctuating metric and perform the calculation for the
specific case of the unit interval.
\end{abstract}
\end{titlepage}

Quantum fluctuations of the geometry of space-time seem to be at the core of the quantization of gravity. In dimensions higher than two, the study of these fluctuations proved to be a formidable task. To study the problem in two dimensions one needs to introduce extra fields. In this work we show that the problem becomes quite simple in a one-dimensional universe. Although our principal reason for studying this case is pedagogical, we believe that our result may be suitable for applications in the statistical mechanics of string-like objects like polymers or magnetic flux lines, or in inhomogeneous optical fibers.

In this work, we present a step-by-step method of calculation of partition
functions for orientable manifolds of fluctuating metric which satisfy a few
basic assumptions. We explicitly perform the calculations for the case of a
specific one-dimensional manifold, the unit interval, where we obtain a
precise answer. Although the method can be easily extended to higher
dimensions, in each case a difficulty has to be overcome, namely the
infinite dimensionality of the space of all configurations of the system. In
the one-dimensional interval this is simply done by dividing this space by
the group that leaves the energy invariant, which in this case is the set of
all diffeomorphisms of the interval that preserve orientation. The result is
a finite-dimensional space where the integration over all accessible states
can be performed.

Let $C${\it \ }denote the set which is the space of all configurations of a
given physical system. The probability of finding the system at a
configuration $c$ at temperature $\beta ^{-1}$ is then given by the
normalized Boltzmann factor 
\begin{equation}
P(c,\beta )=\frac{e^{-\beta E(c)}}{Z(\beta )}, 
\end{equation}
where 
\begin{equation}
Z(\beta )=\sum_{c\in C}e^{-\beta E(c)}=\mathrel{\mathop{\int}\limits_{C}}%
e^{-\beta E(c)}\mu 
\end{equation}
is the partition function. We are here interested in the case where our
physical system is an orientable manifold $M$ and where the different
configurations of $M$ are the different metrics which it can be endowed
with. Thus, $C$ is the space of all possible metrics of $M$ and $\mu $ is a
suitable measure on $C$. 

In general, it is not an easy task to obtain $Z(\beta )$
in a meaningful way. Among the few cases where this is known are the
notorious Polyakov's Bosonic~\cite{Pol1} and Fermionic~\cite{Pol2} strings
that require respectively a 26-component Bosonic field or a 10-component
Fermionic field. In this work we deal with another special case, the unit
interval, where $Z(\beta )$ is calculated without recurring to any extra
fields.

As in any problem in physics, identification of the symmetries of the system
greatly improves the calculations. In this case it is important to find out
whether there exists a group $G$ that acting on $C$ leaves the energy $E$
invariant. The identification of $G$ permits Faddeev-Popov regularization~\cite{Fad};
that is, the functional integral over $C$ in Eq. (2) may have the
volume of the group $G$ factored out. This results in a functional integral
over a somewhat ``smaller '' space, the space of the orbits of $C$ under the
action of $G$. In order to see this let us first review two basic theorems of
analysis~\cite{Spi}:

a) Fubini's theorem states that

$$
\mathrel{\mathop{\int }\limits_{A\times B}}F(x,y)dxdy=%
\mathrel{\mathop{\int }\limits_{B}}\left( \mathrel{\mathop{\int }\limits_{A}}%
F(x,y)dx\right) dy. 
$$

b) change of variables in $\Re^n$ makes

$$
\mathrel{\mathop{\int }\limits_{W}}F(y)dy=\mathrel{\mathop{\int }\limits_{V}}%
F(L(x))\sqrt{\det L^{\prime *}\circ L^{\prime }(x)}dx, 
$$
where $W$ and $V$ are open sets of $\Re^n$ related by a diffeomorphism

$$
\begin{array}{c}
L:V\rightarrow W \\ 
x\mapsto y 
\end{array}
$$
that performs the change of variables, $\circ $ is the usual symbol
indicating the composition of functions and $\det $ stands for determinant.
If the vector spaces $W$ and $V$ have inner products respectively given by $%
<,>$ and $<<,>>$, the adjoint application $L^{*}:W\rightarrow V$ is defined
by $<<Lv,w>>=<v,L^{*}w>$ where $v\in V$ and $w\in W$. Notice that $L^{\prime
*}\circ L^{\prime }(x)$ $:T_x(V)\rightarrow T_x(V)$, where $T_x(V)$ is the
space tangent to $V$ at $x$. This makes $\det L^{\prime *}\circ L^{\prime
}(x)$ independent of the basis of $T_x(V)$.

Fubini's theorem implies that

\begin{equation}
\mathrel{\mathop{\int }\limits_{C}}f(c)\mu =%
\mathrel{\mathop{\int }\limits_{C/G}}\left[%
\mathrel{\mathop{\int }\limits_{[c]}}f(c)\mu _c\right] \mu _{\tilde{g}}, 
\end{equation}
where $c$ is a metric on $C$ and $\tilde{g}$ is the corresponding metric on $%
C/G$, $\left[ c\right] $ denotes the orbit of $c$ under the group $G$; i.e.,
the set of all elements of $C$ obtained from the action of $G$ on the
element $c\in C$.

Now, we want to change the integration over $[c]$ into an integration over $%
G.$ Let 
$$
\begin{array}{c}
A_{c}:G\rightarrow C \\ 
g\mapsto g.c 
\end{array}
$$
be the action of $G$ on $C$. The theorem on the change of variables implies that

\begin{equation}
\mathrel{\mathop{\int }\limits_{[c]}}f(c)\mu _c=%
\mathrel{\mathop{\int }\limits_{G}}f(g.c)FP(c,g)\mu _g,
\end{equation}
where $FP(c,g)$ is the Faddeev-Popov factor

\begin{equation}
FP(c,g)=\sqrt{\det A_{c}^{\prime *}(g)\circ A_{c}^{\prime }(g)}. 
\end{equation}
From now on we will restrict ourselves to the special case where $i$) the
adjoint representation of $G$ is orthogonal and $ii$) $G$ acts on $C$ by
isometries . In terms of the Faddeev-Popov factor this implies that

$$
i)FP(c,g)=FP(c,e) 
$$

$$
ii)FP(c,e)=FP(gc,e), 
$$
respectively, where $e$ is the unit element in $G$. (i) and (ii) allow us to define 
\begin{equation}
FP([c])=FP(c,e)
\end{equation}
Requiring that $f$ be
invariant under $G$ (i.e., $f(gc)=f(c)$ and property ($i$) lead to

\begin{equation}
\mathrel{\mathop{\int }\limits_{[c]}}f(c)\mu _c=%
\mathrel{\mathop{\int }\limits_{G}}f(gc)FP(c,g)\mu _g=f(c)F([c])%
\mathrel{\mathop{\int }\limits_{G}}\mu _g. 
\end{equation}

Here it becomes clear the need of the special conditions $(i)$ and $(ii)$: they
allow us to factor out the volume of the group, a quantity whose method of calculation is not yet known. In fact, to the best of our knowledge,
measures on groups of diffeomorphisms are not known - even in the case of $%
Diff^{+}(I)$, the diffeomorphisms of the unit interval that preserve
orientation, which we will be studying below. On the other hand, Ashtekar
and Lewandowski~\cite{Ash} recently showed that diffeomorphism invariant
theories have well defined measures on the space of connections hinting that
it might be more practical to work with connections instead of metrics.

With $f(c)=e^{-\beta E(c)}$ and using (3) and (7) we define the renormalized
partition function as 
\begin{equation}
Z_{ren}(\beta )=\mathrel{\mathop{\int }\limits_{C/G}}e^{-\beta E(c)}%
FP([c])\mu _{\tilde{g}}. 
\end{equation}
Here $\mu _{\tilde{g}}$ denotes the volume element of $C/G$, which in the case $C=Met(I)$ and $G=Diff^{+}(I)$, that we will be dealing below, is found to be~\cite{Ped}
\begin{equation}
\mu _{\tilde{g}}=c^{*}\left(\frac{2dt}{\sqrt{t}}\right),
\end{equation}
where $c$ denotes the chart that sends $[g]\mapsto l_{g}$ and $*$ indicates the pull-back operation on differential forms.
In this way the volume $\mathrel{\mathop{\int }\limits_{G}}\mu _g$ of the
group $G$ is left out of $Z$. This is justified by the fact that usually one
is only interested in thermodynamic averages computed using the probability
distribution (1); i.e., 
\begin{equation}
\stackrel{\_}{A}=\frac{\mathrel{\mathop{\int }\limits_{C}}A(c)e^{-\beta E(c)}%
}{Z(\beta )}. 
\end{equation}
Since all metrics in $[c]$ have the same energy, they will have the same
weight in the computation of the average. That is, the volume of the group
appears both in the numerator and denominator of (10) cancelling each other.
In this way, by keeping $\mathrel{\mathop{\int }\limits_{G}}\mu _g$ out of
the definition of $Z(\beta )$ we mean that the degeneracies of the energy
function have already been taken into account and for the purpose of
computing thermodynamic averages the space of integration is $C/G$ instead
of $C$. Now, each orbit enters $Z_{ren}(\beta )$ with a weight precisely
defined by the Faddeev-Popov factor.

In what follows we restrict ourselves to the case where $M$ is the unit
interval $\left[ 0,1\right] $. In this case the space of the orbits of $C$
is one-dimensional permiting a straightforward evaluation of $Z$. Although
this calculation may be extended to other manifolds, feasibility of the
method depends on one being able to reduce the space where functional
integration should be performed to a finite-dimensional one.


Our configuration space will be the set of all metrics of $I=[0,1]$; that
is, 
\begin{equation}
C=\left\{ \alpha dt\otimes dt|\alpha :I\rightarrow \Re^{+}\right\} =Met(I).
\end{equation}
The energy $E(g)$ is in fact a function of the length (volume) of the
interval $l_g=\mathrel{\mathop{\int }\limits_{I}}v_g=$, where $v_g$ is the
volume element corresponding to the metric $g$.

The symmetry group of $E$ is then the group $G=Diff^{+}(I)$, the group of the diffeomorphisms of the interval $I$ that preserve orientation, since this group keeps $l_g$ invariant. In other words,

$$
E(f^{*}g)=E(g),\forall f\in G. 
$$
Here $f^{*}g$ denotes the pull-back of $g$ by the diffeomorphism $f$. G satisfies conditions (i) and (ii) of section 1 enabling us to use Eq. (8) to compute the partition function. We need now to compute the Faddeev-Popov factor, Eq. (5).

We start with
\begin{equation}
A'^{*}_{g}A'_{g}: \Gamma_{0}(TI) \longrightarrow  \Gamma_{0}(TI)= T_{id}(Diff^{+}(I)),
\end{equation}
where $\Gamma_{0}(TI)$ is the set of sections of the tangent bundle over $I$ that vanish at the boundary of $I$ and $T_{id}(Diff^{+}(I))$ denotes the tangent space to $Diff^{+}(I)$ at the identity $id$. In this case we find
\begin{equation}
A'^{*}_{g}A'_{g}(\varphi e)=4(-\triangle_{g}\varphi)e,
\end{equation}
where $e$ is the global orthonormal oriented reference frame for the metric $g$.
After a few more steps and using Eq. (22) from the appendix and the well-known fact~\cite{Zet} about the ordinary Riemann's $\zeta$-function that $\zeta (0)=-1/2$ and $\zeta '(0)=-\frac{1}{2}\ln 2\pi$, we get
\begin{equation}
det A'^{*}_{g}A'_{g}=l_{g}.
\end{equation}
It follows that
\begin{equation}
FP([g])=l_{g}^{1/2} .
\end{equation}
Using now Eqs. (15) and (9) in (8) we end up with
\begin{equation}
Z_{ren}(\beta)=2\int_{0}^{\infty}e^{-\beta E(l_g)}dl_{g},
\end{equation}
which is the main result of this work. 

Now, if we choose the energy to be quadratic in $l_g$; i.e., $E(l_{g})=l_g^{2}$ then we find
\begin{equation}
Z(\beta)=\sqrt \frac{\pi}{\beta}.
\end{equation}
This partition function gives 
\begin{equation}
\bar{E}=\frac{1}{2}k_{B}T
\end{equation}
for the mean energy. Mathematically, it is no surprise that we obtain the one-dimensional classical ideal gas solution since there is a one-to-one correspondence between the length $l_g$ and the atomic velocity. Physically, it is intriguing that there should be any relation between fluctuating metrics in the interval and the classical ideal gas. It would be interesting (and perhaps quite useful) to find out if this also happens in higher dimensions.

{\bf Appendix: $\zeta$-function Regularization}
\\
The determinant of a self-adjoint operator $\hat P$, acting on a finite-dimensional vector space $V$, is the product of its eigenvalues 
${p_i}$: 
\begin{equation}
det \hat P=\mathrel{\mathop{\prod}\limits_{i}}p_i . 
\end{equation}
Assuming $0<p_{1}<p_{2}<...$ we write Riemann's generalized $\zeta$-function
associated to $\hat P$ as 
\begin{equation}
\zeta_{\hat P}(s)=\mathrel{\mathop{\sum}\limits_{i}}p_{i}^{-s}. 
\end{equation}
The determinant of $\hat P$ can be thus rewriten as 
\begin{equation}
det \hat P=\exp (-\zeta_{\hat P}^{\prime}(0)) 
\end{equation}
and we also have that $dim V=\zeta_{\hat P}(0)$. A straightforward
calculation yields the following useful result: 
\begin{equation}
det \beta \hat P=\beta^{\zeta_{\hat P}(0)} det \hat P . 
\end{equation}
This last equation then makes sense for an elliptic self-adjoint operator $\hat P$ acting on the section $\Gamma (F)$ of a fiber bundle $F$ over a compact base, which in this particular case is the unit interval $I$. This is so, because $\zeta_{\hat P}(s)$ is a meromorphic function, holomorphic at $s=0$.
\noindent

{\bf Acknowledgment}\\ 
\noindent
This work was partially supported by CNPq.

\end{document}